\documentclass[sigconf]{acmart}
\usepackage{soul}
\usepackage{url}
\usepackage{graphicx}
\usepackage{amsmath}
\usepackage{amsfonts}
\usepackage{multirow}
\usepackage{amsthm}
\usepackage{booktabs}
\usepackage{epstopdf}
\usepackage{xcolor}
\usepackage[labelformat=simple]{subcaption}
\usepackage{caption}
\usepackage[ruled,linesnumbered]{algorithm2e}
\usepackage{setspace}

\theoremstyle{definition}
\newtheorem{myDef}{Definition}[section]
\newtheorem{example}{Example}

\newcommand{\tabincell}[2]{\begin{tabular}{@{}#1@{}}#2\end{tabular}}

\graphicspath{{images/}}

\AtBeginDocument{%
  \providecommand\BibTeX{{%
    \normalfont B\kern-0.5em{\scshape i\kern-0.25em b}\kern-0.8em\TeX}}}

\copyrightyear{2021}
\acmYear{2021}
\setcopyright{acmcopyright}
\acmConference[CIKM '21]{Proceedings of the 30th ACM International Conference on Information and Knowledge Management}{November 1--5, 2021}{Virtual Event, QLD, Australia}
\acmBooktitle{Proceedings of the 30th ACM International Conference on Information and Knowledge Management (CIKM '21), November 1--5, 2021, Virtual Event, QLD, Australia}
\acmPrice{15.00}
\acmDOI{10.1145/3459637.3482250}
\acmISBN{978-1-4503-8446-9/21/11}

\begin{document}
\fancyhead{}
\title{Detecting Communities from Heterogeneous Graphs: A Context Path-based Graph Neural Network Model}

\author{Linhao Luo}
\affiliation{%
  \institution{Harbin Institute of Technology, Shenzhen}
    \country{China}
}
\email{luolinhao@stu.hit.edu.cn}

\author{Yixiang Fang}
\authornote{\ Corresponding authors}
\affiliation{%
  \institution{The Chinese University of Hong Kong, Shenzhen}
  \country{China}}
\email{fangyixiang@cuhk.edu.cn}

\author{Xin Cao}
\affiliation{%
  \institution{The University of New South Wales}
  \country{Australia}}
\email{xin.cao@unsw.edu.au}

\author{Xiaofeng Zhang}
\authornotemark[1]
\affiliation{%
  \institution{Harbin Institute of Technology, Shenzhen}
    \country{China}
}
\email{zhangxiaofeng@hit.edu.cn}

\author{Wenjie Zhang}
\affiliation{%
  \institution{The University of New South Wales}
  \country{Australia}}
\email{wenjie.zhang@unsw.edu.au}


\begin{abstract}

Community detection, aiming to group the graph nodes into clusters with dense inner-connection, is a fundamental graph mining task. Recently, it has been studied on the heterogeneous graph, which contains multiple types of nodes and edges, posing great challenges for modeling the high-order relationship between nodes. With the surge of graph embedding mechanism, it has also been adopted to community detection. A remarkable group of works use the meta-path to capture the high-order relationship between nodes and embed them into nodes' embedding to facilitate community detection. However, defining meaningful meta-paths requires much domain knowledge, which largely limits their applications, especially on schema-rich heterogeneous graphs like knowledge graphs.
To alleviate this issue, in this paper, we propose to exploit the context path to capture the high-order relationship between nodes, and build a Context Path-based Graph Neural Network (CP-GNN) model. It recursively embeds the high-order relationship between nodes into the node embedding with attention mechanisms to discriminate the importance of different relationships. By maximizing the expectation of the co-occurrence of nodes connected by context paths, the model can learn the nodes' embeddings that both well preserve the high-order relationship between nodes and are helpful for community detection. Extensive experimental results on four real-world datasets show that CP-GNN outperforms the state-of-the-art community detection methods \footnote{Code and data are available at: https://github.com/RManLuo/CP-GNN}.

\end{abstract}




\maketitle

\section{Introduction}
\label{sec:intro}

\begin{figure}[htbp]
    \centering
    \hspace*{-.5cm}
    \includegraphics[width=1.08\columnwidth]{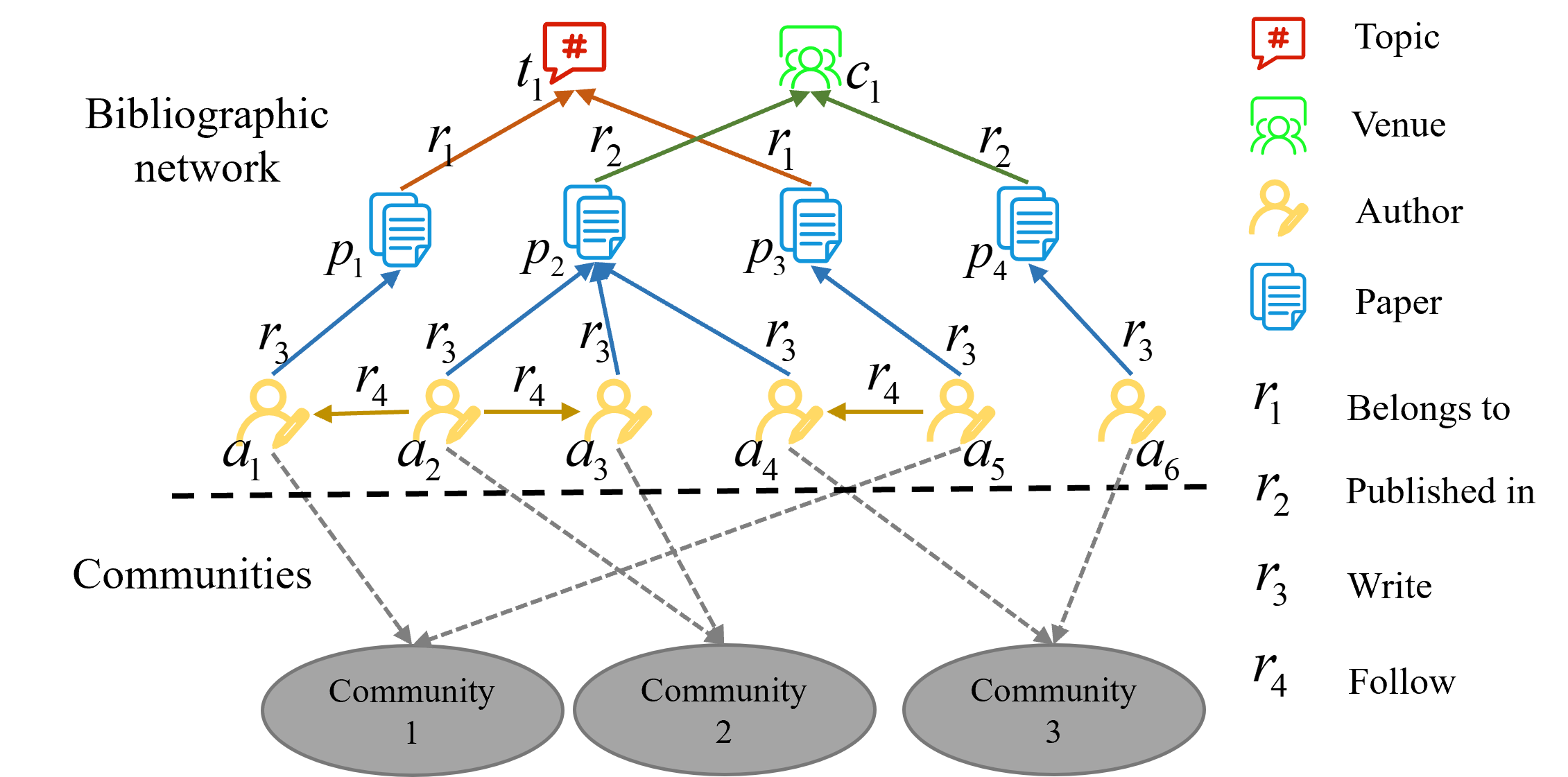}
    \caption{Detecting communities from a bibliographic graph.}
    \label{fig:HINCommunityDetection}
\end{figure}

As a fundamental topic in network science, community detection, aiming to group the graph nodes into clusters with dense inner-connection, has been studied for decades and found various real-world applications, 
such as recommendation (e.g., \cite{satuluri2020simclusters, luo2020motif}), anomaly detection (e.g., \cite{wang2016botnet}), and scientific discipline discovery (e.g., \cite{zhou2009graph}).
Most existing works of community detection (e.g., \cite{wakita2007finding,chen2019supervised,fang2020survey}) mainly focus on detecting communities from homogeneous network that contains the same type of nodes.
These solutions, however, may not work well on many real-world graphs that are with multiple node types and edge types, which are also called {\it heterogeneous graphs}, and they are prevalent in many real-world applications, including bibliographic networks, social media, and knowledge graphs. For example, Figure~\ref{fig:HINCommunityDetection} depicts a bibliographic network with four types of nodes, i.e., {\it paper}, {\it author}, {\it venue}, and {\it topic}, and four relations (edge types) among them.
The existing works of community detection on heterogeneous graphs can generally be classified into two groups; the first group \cite{sun2009ranking} focuses on detecting clusters, each of which contains objects with multiple types, and the second group \cite{boden2014density} aims to generate clusters of nodes with a specific type. In this paper, we follow the second group and aim to cluster nodes such that nodes in the same cluster have strong relationships.

Detecting communities from heterogeneous graphs is more challenging than that on homogeneous graphs, since the multiple types of nodes and edges carry more abundant semantic information.
In Figure~\ref{fig:HINCommunityDetection}, for example, the authors $a_1$ and $a_5$ should belong to the same community (Community 1), but they are not directly connected in the graph, making them hard to be grouped into the same cluster if only direct relations (e.g., ``follow'') are considered). However, $a_1$ and $a_5$ have written papers $p_1$ and $p_3$ respectively, which share the same topic $t_1$. As a result, they can be grouped into the same cluster if we consider their high-order relationship, or the relationship that cannot be captured by directed links.

To capture the high-order relationship above, several efforts have been made (e.g., \cite{sun2013pathselclus,fang2020effective}), but most of them rely on some pre-defined meta-paths~\cite{sun2011pathsim}, which reveals the latent high-order relationships. For example, the path \textit{author-paper-topic-paper-author} can model the relationship we showed above. To further capture and represent the high-order relations, a few works \cite{dong2017metapath2vec,zheng2019heterogeneous} integrate the meta-path oriented graph embedding mechanism with community detection.
However, the problem of these methods is that their performance highly depends on the quality of the pre-defined meta-paths, which need to be selected by domain experts.
Moreover, the number of meta-paths increases exponentially with the path length, meaning that it is almost infeasible to find all meaningful meta-paths to capture the high-order relationships.
Furthermore, different meta-paths contribute differently to the community detection, which imposes great challenges for distinguishing their importance.

In this paper, we propose a novel model, called the \underline{C}ontext \underline{P}ath-based \underline{G}raph \underline{N}eural \underline{N}etwork (CP-GNN), for detecting communities with nodes of the same target type in the heterogeneous graph. Here, the target type is also called the {\it primary type}, while the other types of nodes are called auxiliary types. In this model, we adopt the concept of ``context path'' \cite{barman2019k}, which links two primary type nodes via a sequence of edges with some auxiliary type nodes. It can not only well capture the high-order relationship, but also avoid requiring customized meta-paths selected by domain experts.
We first introduce the \textit{context path probability} that is the probability that two nodes are connected by a context path.
Then, we propose a novel objective function for learning node embeddings unsupervisedly, by maximizing the expectation of the co-occurrence of context neighbors (nodes connected by context paths), which exploits both the structure and the high-order context relationships among nodes.

To learn the embeddings, instead of exhaustedly enumerating all the context paths, we employ the graph neural network model to recursively embed the context path information between nodes into the node embeddings.
We further propose the length-wise and relation-wise attention mechanisms to discriminate the importance of different context paths that preserve different high-order relationships.
Thus, our model not only avoids customizing the meta-paths, but also well captures the high-order relationships between nodes that are preserved by the context paths with different importance.
Finally, the learned embeddings from the neural network are directly used for community detection.

In summary, our principal contributions are as follows:
\begin{itemize}
    \item We adopt the context path to capture the high-order relationship information and introduce the context path probability to model the learning objective function.
    \item We propose a novel neural network model CP-GNN, which can capture the rich high-order relationship for learning the node embeddings unsupervisedly.
    \item We conduct extensive experiments on four real datasets, which demonstrate the superior performance of our CP-GNN model over the state-of-the-art methods. And the visualization experiments show the CP-GNN can capture high-order relationships with different importance.
\end{itemize}

\section{Related Work}
\label{sec:related}

In this section, we review three representative groups of existing works on the topic of network community detection.

\noindent$\bullet$ \textbf{Conventional community detection.}
Community detection has attracted a lot of research attention \cite{newman2004finding,fortunato2010community,fang2020,chen2020efficient}. The early works often exploit the local link structure to group the the vertices into different clusters \cite{wakita2007finding,sales2007extracting}. More related works can be found in these survey papers \cite{malliaros2013clustering,fang2020survey}.

However, most of these methods focus on homogeneous graphs. Recently, some works have studied the community detection task on heterogeneous graph \cite{shi2016survey,moscato2021survey}. Reference \cite{cai2005mining} proposes a method that can learn an optimal linear combination of the relations in heterogeneous graph. Then by adopting MinCut-based and Regression-based algorithms, it can achieve a better performance on community detection. Reference \cite{qi2012clustering} models the structure and content of heterogeneous graph with outlier links. HeProjI \cite{shi2014ranking} projects a heterogeneous graph into a sequence of sub-networks and conducts community detection. TCSC \cite{boden2014density} considers both the graph connection and vertex attributes to detect clusters. AGGMMR \cite{zhe2019community} proposes a framework to perform community detection utilizing both the attributes and topological information through a greedy modularity maximization model. Reference \cite{sun2013pathselclus,fang2020effective} adopt the meta-path to capture the high-order relationships between nodes for detecting communities in heterogeneous graphs. Nevertheless, as aforementioned, the meta-paths need to be carefully selected by domain experts, which imposes a great limitation for their applications.

\noindent$\bullet$ \textbf{Graph embedding for community detection.} Recently, with the surge of graph embedding methods \cite{grover2016node2vec,dong2017metapath2vec}, many researchers focus on addressing the community detection problem with the help of graph embedding \cite{tian2014learning,xie2016unsupervised,yang2016modularity}. Cavallari et al \cite{cavallari2017learning} intergates the node embedding and clustering together to conduct community detection by optimizing the first-order and second-order neighbors' loss, high-order loss, and clustering loss. CDE \cite{li2018community} proposes a novel embedding based method. It embeds the inherent community structures into structure embeddings via known community memberships. Then based on the node attributes and community structures embeddings, it formulates the community detection as a matrix factorization optimization problem. NEC \cite{sun2020network} proposes an algorithm to learn graph embedding for community detection in heterogeneous graphs, which learns graph structure-based representations and clustering-oriented representations together. Then it adopts the K-means to perform the community detection.

\noindent$\bullet$ \textbf{GNN-based community detection.}
Many deep learning-based community detection methods are also developed \cite{jin2021survey}.
As one of the most widely used deep learning techniques, the Graph Neural Network (GNN) \cite{kipf2016semi} has also shown great power in community detection \cite{chen2019supervised,zhang2020community}. LGNN \cite{chen2019supervised} is a graph neural network model which exploits edges' adjacency information of the graph for community detection. MRFasGCN \cite{zhang2020community} proposes a Markov random field enhanced GNN to group nodes into different communities. However, most of them do not consider complex relationships, which leads them inadequate to fuse enough relationships for heterogeneous graph community detection. Recently, HTGCN \cite{zheng2019heterogeneous} shows a temporal graph neural network to perform the community detection on temporal heterogeneous graph. It considers both the temporal and heterogeneous information of the graph to increase the performance. Despite the existing success, most GNN-based approaches \cite{chen2019supervised,wang2019heterogeneous} regard the community detection as a supervised node classification task, which predicts the target community for each node. However, the ground truths of the community are not always available, making them inapplicable in this case. Thus, it is desirable to develop fully unsupervised GNN-based community detection methods.

\section{Preliminaries}
\label{sec:pre}


\begin{myDef}[\textbf{Heterogeneous graph  \cite{sun2011pathsim}}]
The heterogeneous graph is defined as a graph $G=(\mathcal{V},\mathcal{E},\mathcal{A}, \mathcal{R})$ with a node mapping function $\phi(v): \mathcal{V}\to \mathcal{A}$ and an edge mapping function $\psi(e):\mathcal{E}\to \mathcal{R}$,
where $|\mathcal{A}| + |\mathcal{R}| > 2$, each node $v\in\mathcal V$ belongs to a node type $\phi(v)\in\mathcal A$, and each edge $e\in\mathcal E$ belongs to an edge type (also called relation) $\psi(e)\in\mathcal R$.
\end{myDef}




\begin{myDef}[\textbf{Primary type and auxiliary type}]
\label{def:ptype}
As aforementioned, for the purpose of community detection, usually only one node type is targeted by the task, and we call it the {\it primary type}, denoted by $P$. The nodes with type $P$ are called primary nodes.
The other node types are called auxiliary types, which constitute a set of types $\mathcal A'$.
Note that technically, any node type can be regarded as the primary type.
\end{myDef}

\begin{myDef}[\textbf{Primary graph and auxiliary graph}]
Given a heterogeneous graph $G$, the primary graph is a subgraph of $G$, denoted by $G_P$=$(\mathcal{V}_P, \mathcal{E}_P)$ where each node $v\in \mathcal{V}_P$ is of the primary node type $P$ and each edge $e \in \mathcal{E}_P \subseteq \{\mathcal{V}_P \times \mathcal{V}_P\}$. Similarly, the auxiliary graph is also a subgraph of $G$ with nodes of a specific type $A$, denoted by $G_A=(\mathcal{V}_A, \mathcal{E}_A)$, where for each $v \in \mathcal{V}_A$, $\phi(v) = A \in\mathcal A'$ and each edge $e\in\mathcal{E}_A\subseteq \{\mathcal{V}_A \times \mathcal{V}_A\}$.
\end{myDef}

\begin{example}
    As the heterogeneous graph shown in Figure \ref{fig:HINCommunityDetection}, the ``Author'' can be defined as the primary type, and the remaining node types are treated as auxiliary types. Similarly, the ``Paper'' can also be chosen as the primary type if necessary. Meanwhile, the ``Author'', ``Topic'', and ``Venue'' nodes and their relations will form the auxiliary graphs.
\end{example}



\begin{myDef}[\textbf{Context path and context neighbors \cite{barman2019k}}]
Given a heterogeneous graph $G$, a {\it context path} is a path connecting two nodes $v_i$ and $v_j$ in the primary graph $G_P$, denoted by $\rho^K=\left \langle v_i, R^K, v_j \right \rangle$, where $R^K$ is any path connecting $v_i$ and $v_j$ that contains only $K$ ($K\geq0$) nodes in auxiliary graphs, and it is also called the {\it context edge sequence}.
The length of a context path is $K$ (when $K$=0, $R^K$=$\emptyset$).
We say two nodes are \textit{context neighbors} if they are connected by a context-path.
\end{myDef}

\begin{figure}[htbp]
\centering
    \includegraphics[trim={0 0 0 4},width=0.55\columnwidth,clip]{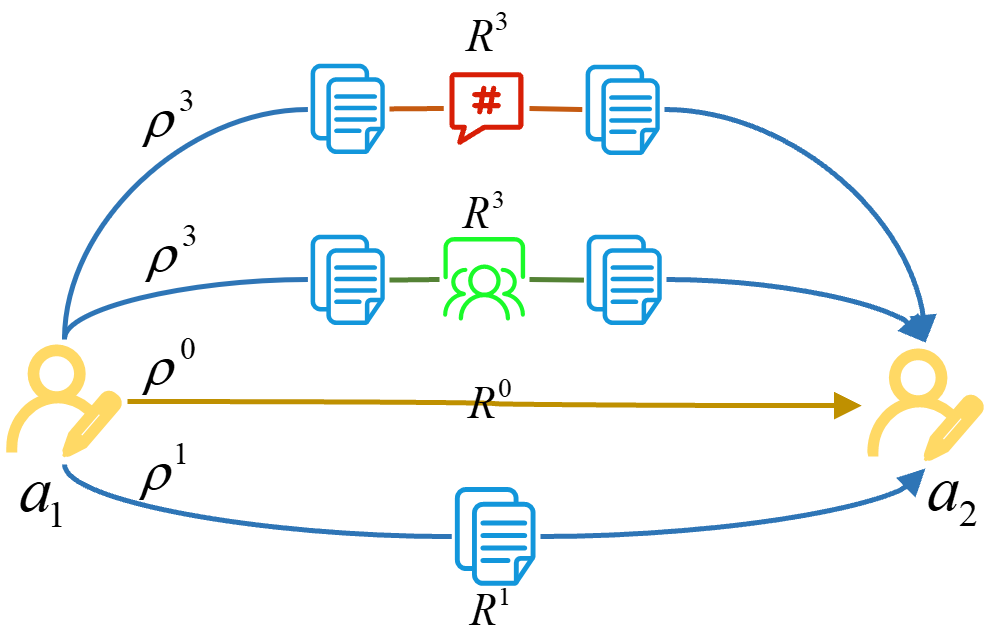}
\caption{Context paths between $a_1$ and $a_2$.}
\label{fig:ContextPath}
\end{figure}

\begin{example}
Figure \ref{fig:ContextPath} depicts four possible context paths $\rho^*$ with different lengths that can connect authors $a_1$ and $a_2$, where $R^*$ denotes the auxiliary nodes that constitute the path.
\end{example}

\noindent\textbf{Difference between context path and meta-path.} Intuitively, the different high-order semantic relationships revealed from the context path come from the different auxiliary nodes. The purpose of the meta-path is to manually define the combination of the auxiliary nodes in the context path. However, the number of the combinations explodes exponentially when the node types and order size increase. Thus, the context path relaxes the restriction of the auxiliary nodes. Given a length $K$, the context path contains all the possible $K$-order relationships. For example, in Figure \ref{fig:ContextPath}, two meta-paths \textit{Author-Paper-Topic-Paper-Author} and \textit{Author-Paper-Venue-Paper-Author} can both be represented by a 3-length context path, and we propose the CP-GNN to futher differentiate them.

Besides, specifying an integer $K$ is much easier than defining the meta-path, because the number of meta-paths of different node/edge types grows exponentially as the meta-path length increases, while the choices of $K$ are rather limited since the average length of the shortest path between two nodes in real-world networks is between 4 to 6, according to \cite{ye2010distance}, thus the $K$ can be determinated empirically or with the help of the proposed context path length attention mechanism.

\noindent\textbf{Problem definition.} Given a heterogeneous graph $G$, our goal is to learn a good primary node embedding $Z_P \in \mathbb{R}^{N\times d}$ where $N$ denotes the number of primary type nodes and $d$ is the embedding dimension, such that they can be used to group the nodes into a set of communities $\mathcal{C}=\{1,\cdots,C\}$ with strong inner-connection.



\begin{figure*}[htbp]
\centering
\includegraphics[width=1\textwidth]{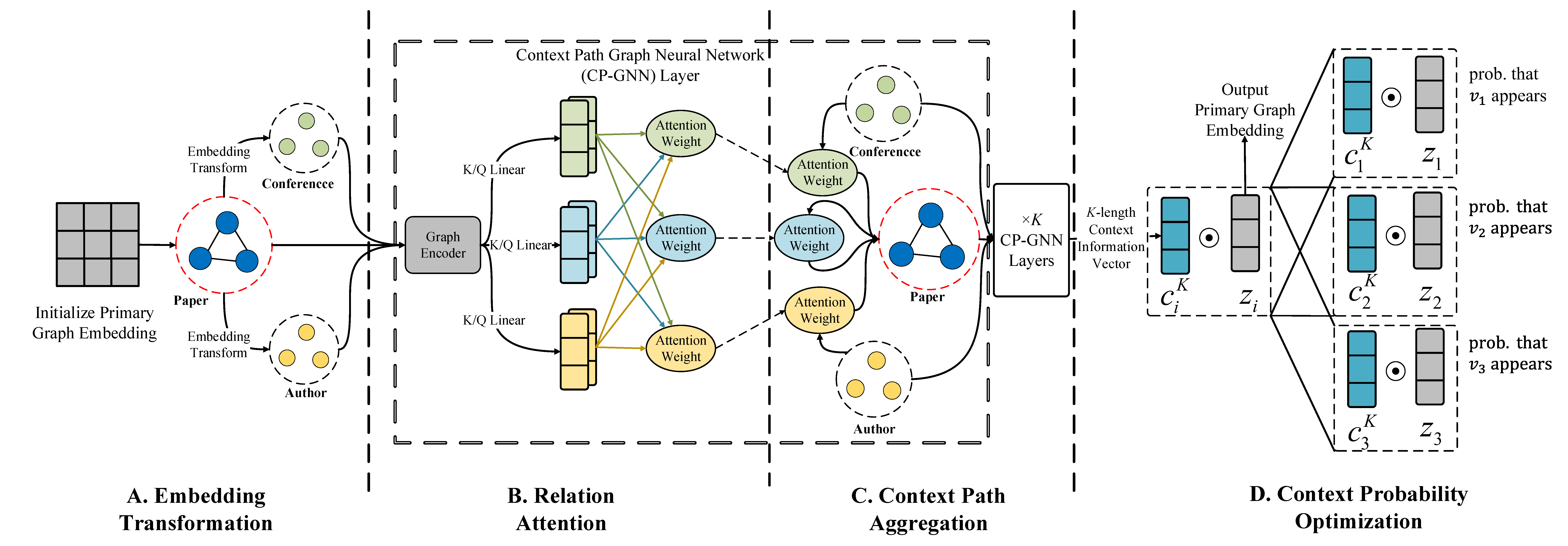}
\caption{The overall framework of the CP-GNN (A: The embedding of each auxiliary graph is transformed from the primary graph. B: The relation attentions are calculated from the corresponding graph representations. C: The context information vector is generated from the CP-GNN. D: The embedding of the primary graph is optimized with the context probability).}
\label{fig:framework}
\end{figure*}

\section{Our CP-GNN Approach}
\label{sec:CP-GNN}

In this section, we present the Context Path-based Graph Neural Network (CP-GNN) model for learning node representations that well preserve the high-order relationship between nodes, and the overall framework is depicted in Figure \ref{fig:framework}.
Given a heterogeneous graph (composed of a primary graph and auxiliary graphs), for each node in the primary graph $G_P$, we first extract all its context neighbors by using the context paths whose lengths range from 0 to $K$.
Then, we learn the representations of nodes in $G_P$ by training the CP-GNN model, in which the objective function is to maximize the probability of having the context neighbors for each node in $G_P$.
In the following, we first introduce the objective function and then discuss the details of the CP-GNN model.

\subsection{Objective Function}
\label{sec:objective}

In this section, we first introduce the context path probability, which is defined as the probability that two nodes $v_i$ and $v_j$ in $G_P$ are context neighbors. Specifically, given a context path $\rho^K$=$\left\langle v_i,R^K,v_j\right\rangle$, the context path probability is
\begin{equation}
    \setlength\abovedisplayskip{3pt}
    \setlength\belowdisplayskip{3pt}
     p(v_j|v_i, R^K; \theta),
    \label{eq:contextprobability}
\end{equation}
\noindent where $\theta$ is the parameters for computing the probability. 

In our model, to learn effective node representations in $G_P$, a maximum length $K$ is firstly given. Then, for each length $k\in [0,K]$, we aim to maximize the co-occurrence probability of all nodes in $G_P$ and their context neighbors w.r.t. all the $k$-length context paths. The objective function can be written as
\begin{equation}
    \setlength\abovedisplayskip{1pt}
    \setlength\belowdisplayskip{1pt}
    arg \mathop{max}\limits_{\theta} \mathcal{O(\theta)} = \sum\limits_{v_i\in G_P} \sum\limits_{k=0}^{K} \sum\limits_{v_j\in N_P^{k}(v_i)} \sum\limits_{\rho^k \in \mathbb{P}^k} log p(v_j|v_i,R^k;\theta),
\end{equation}
\noindent where $N_P^{k}(v_i)$ is the set of $k$-length context neighbors of $v_i$ in $G_P$ and $\mathbb{P}^k$ is a set of all $k$-length context paths connecting $v_i$ and $v_j$.

Intrinsically, Breadth-first search (BFS) is the easiest way to get all the context paths between nodes. However, the number of context paths increases exponentially with the path length, which makes it impossible to traverse all the context paths, and the walk-based methods (e.g., Node2vec \cite{grover2016node2vec}, Metapath2vec \cite{zhang2018metagraph2vec}) are also computational heavily and cannot fully excavate the relationships.

To address this issue, in our CP-GNN model, we adopt graph neural network to recursively embed the high-order relationship of each node into a context information vector $c^k$ to represent the relation information of all context paths with length $k$, which takes linear time complexity cost. Many previous researches have already adopted the GNN to capture the structure and path information in graph \cite{xu2018powerful,wang2020self}. The GNN message passing is essentially a simulation of BFS, which exhibits the ability to capture paths between nodes \cite{you2021identity}. 

After $k$ times message passing, CP-GNN can embed all the $k$-length context paths into the context vector $c^k$. Then, the probability of two context neighbor nodes $v_i$ and $v_j$ connected by all the possible $k$-length context paths can be approximated by their respective context information vectors $c_i^k$ and $c_j^k$, written as $p(v_j|v_i,c_i^k,c_j^k;\theta)$, which could be calculated using a softmax function:
\begin{equation}\label{eq:cprob}
    \setlength\abovedisplayskip{1pt}
    \setlength\belowdisplayskip{1pt}
\begin{aligned}
    p(v_j|v_i,c_i^k,c_j^k; \theta) &= \frac{exp\left(\varphi(z_i,c_i^k,c_j^k,z_j)\right)}{\sum\limits_{v_x\in G_P}exp\left(\varphi(z_i,c_i^k,c_{x}^k,z_{x})\right)}\\
    \varphi(z_i,c_i^k,c_j^k,z_j)  &= \sigma\left((z_i\odot c_i^k)^\top (z_j\odot c_j^k)\right),
\end{aligned}
\end{equation}
where $z_i$ and $z_j$ are the node embeddings of $v_i$ and $v_j$ we want to learn, $c_i^k$ and $c_j^k$ denote the context information vectors, $\odot$ denotes the element-wise vector product operation, and $\sigma(\cdot)$ denotes the sigmoid function. 

Thus, the objective function could be futher simplified as
\begin{equation}
    \setlength\abovedisplayskip{1pt}
    \setlength\belowdisplayskip{1pt}
        arg \mathop{max}\limits_{\theta} \mathcal{O(\theta)} = \sum\limits_{v_i\in G_P} \sum\limits_{k=0}^{K} \alpha_k\sum\limits_{v_j\in N_P^{k}(v_i)} log p(v_j|v_i,c_i^k,c_j^k;\theta),
\end{equation}
where $N_P^{k}(v_i)$ is the set of $k$-length context neighbors of $v_i$ in $G_P$.


To differentiate the importance of context paths with different lengths from 0 to $K$, we propose a \textbf{Context Path Length Attention} mechanism to assign attention weights for different path lengths. For the $k$-length context path, we use $\alpha_k$ to denote its attention weight, which indicates the importance of the $k$-length context paths and $\alpha_k \in (0,1]$.
Inspired by a multi-task leaning method \cite{kendall2018multi}, we adopt the similar way to optimize $\alpha_k$ during the training. By considering the negative sampling technique, the final model objective function is converted to the following loss function:
\begin{equation}
    \setlength\abovedisplayskip{1pt}
    \setlength\belowdisplayskip{1pt}
\begin{split}
        \mathcal{L } =
    \sum\limits_{v_i\in G_P}
    \bigg(
        \sum\limits_{k=0}^K -\alpha_k
        \Big(
            \sum\limits_{v_j\in N_P^{k}(v_i)} log\varphi(z_i,c_i^k,c_j^k,z_j) + \\
            \sum\limits_{v_{x}\in N_P^{k-}(v_i)}log-\varphi(z_i,c_i^k,c_{x}^k,z_{x})
        \Big)
        - log\alpha_k
    \bigg),
\end{split}
\end{equation}
where $N_P^{k-}$ is the set of negative context neighbors of $v_i$ and $\log \alpha_k$ is a penalty to prevent $\alpha_k$ from over small. Currently, the computation complexity of CP-GNN is $O(K \times |V_P| \times (n^++n^-))$ that grows linearly with $K$, where $K$ is the defined maximal context path length, $|V_P|$ is the number of primary nodes, and $n^+,n^-$ are the numbers of the positive and negative sampling neighbors.

Thus we can optimize the parameters with gradient descent written as $\theta\leftarrow\theta- \gamma\bigtriangledown_{\theta}\mathcal L(\theta)$, where $\gamma$ is the learning rate.


\subsection{Details of CP-GNN Model}
\label{sec:details}

Our CP-GNN model aims to capture the context information and generate the context information vectors for optimizting the final node embedding. 
CP-GNN has two major components: {\it Embedding Transformation} for transforming the node embedding from the primary graph to auxiliary graphs, and {\it CP-GNN Layer} for embedding the high-order relationship of each node into a context information vector.
%

\subsubsection{Embedding Transformation}
\label{sec:embtransform}
This component is used to transform the node embedding from the primary graph to the auxiliary graphs along the edges. In this way, we only learn the embedding of nodes in the primary graph thus reducing the parameters to be learned, as we only need to learn the transformation weight matrix.

We also observe that this can achieve even better performance than directly learning the node embeddings for auxiliary graphs in the experiments as shown in Section~\ref{sec:parameter}. Because we can learn the representations of the primary graph by exploiting the information from the auxiliary graphs by embedding transformation. It can generate the representations of primary graph under different contexts (auxiliary graphs), which is essential in heterogeneous graph representation and community detection \cite{epasto2019single,liu2019single,park2020unsupervised}.

The embedding transformation function $\mathcal{T}_{ST}(\cdot)$ from one graph $G_S$ with node type $S$ to another graph $G_T$ with node type $T$ is
\begin{equation}
    Z_T = \sigma(A_{ST}{Z_S}W + B),
    \label{eq:transfunc}
\end{equation}
where $Z_S$ and $Z_T$ respectively denote the embeddings of nodes in $G_S$ and $G_T$, $A_{ST}$ is a bipartite adjacency matrix between $G_S$ and $G_T$, $W_{ST}$ is the transformation weight matrix, and $\sigma(\cdot)$ is the non-linear activation function such as ReLU. The ReLU can be seen as a ``Mask'' for filtering unnecessary embedding features (values less than 0) during the transformation \cite{yang2020multisage}.


\subsubsection{Context Path Graph Neural Network Layer}

This component recursively captures the high-order relationship from the graph by repeating the Relation Attention and Context Path Aggregation operations.

\textbf{Relation Attention} aims to calculate the attention score of each relation, so that the contributions of different relations are well differentiated.
We first use a graph encoder to encode each graph to a summary vector $h$. After that, the attention score of each relation is calculated based on the graph summary vectors.


To enhance the model robustness, the graph encoder contains a node dropout mechanism which randomly drops nodes from the original graph. Then an averaging operation is adopted to calculate the global graph representation $h$.
Although there exist several techniques to generate the graph summary vector $h$, the simple averaging operation demonstrates superior performance \cite{ren2019heterogeneous}, and thus $h$ is calculated as
\begin{equation}
    \setlength\abovedisplayskip{1pt}
    \setlength\belowdisplayskip{1pt}
    \begin{split}
        G' &= NodeDropout(G)\\
        h  &= Mean(C'),
    \end{split}
\end{equation}
where $C'$ contains the context information vectors of all the nodes in the graph $G'$.

After calculating the $h$, for the $l$-th CP-GNN layer, we calculate the $h$-head attention score $\alpha_{ST}^{h,l}$ for each relation $r_{ST} \in \mathcal{R}$ by
\begin{equation}\label{eq:rlagg}
    \setlength\abovedisplayskip{1pt}
    \setlength\belowdisplayskip{1pt}
    \begin{split}
        & h_S^l = GraphEncoder(C_S^{l-1})\\
        & h_T^l = GraphEncoder(C_T^{l-1})\\
        & \alpha_{ST}^{h,l} = \mathop{Softmax}\limits_{S\in \mathcal{A}} \frac{Q^h(h_T^l)^\top K^h(h_S^l)}{\sqrt{d}}\\
        & Q^h(h_T^l) = QLinear_{T}^h(h_T^l)\\
        & K^h(h_S^l) = KLinear_{S}^h(h_S^l),\\
    \end{split}
\end{equation}
where $S$ and $T$ respectively denote the source and target node types in the relation $r_{ST}$, $h_S^l$ and $h_T^l$ denote the graph summary vector of $G_S$ and $G_T$ at layer $l$ respectively, $C_S^{l-1}$ and $C_T^{l-1}$ are the context information vectors at the ($l-1$)-th layer, $QLinear$ and $KLinear$ are the linear projection functions that project the graph summary vectors to a Query vector and a Key vector. We want to learn more diverse importance of the relations, thus we adopt total $H$ different heads of Relation Attention with their own parameters to be learned during the training. $\alpha_{ST}^{h,l}$ is the attention weight in head $h$ at layer $l$ for the relation $r_{ST}$.

\begin{figure}[htbp]
    \centering
    \includegraphics[width=0.9\columnwidth]{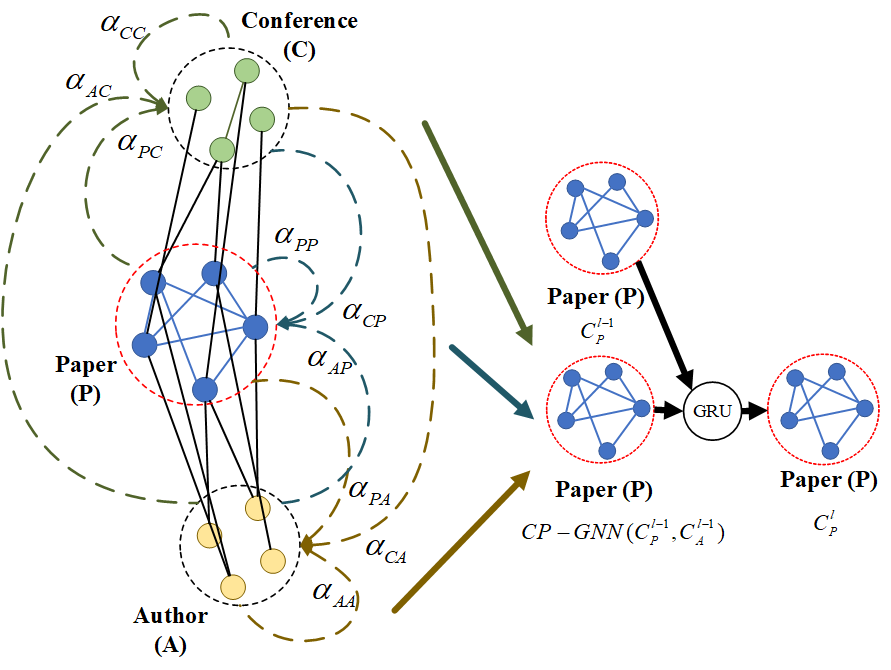}
    \caption{The Context Path Aggregation component.}
    \label{fig:relationaggregation}
\end{figure}


\textbf{Context Path Aggregation} aims to aggregate the information along relations to generate the context information vectors for all nodes. As discussed in Section \ref{sec:objective}, it is infeasible to enumerate all the context paths with lengths at most $K$, so we propose to use the context information vectors to approximately compute the probability that two context neighbors are connected by a context path. After calculating the scores of different relationships, we aggregate the information for a node $v_i$ of type $T$ from its one-hop neighbors by adopting the widely used GNN aggregation method. The procedure of this component is shown in Figure \ref{fig:relationaggregation}.

Assume that at layer $l$, we are going to obtain the context information vector $c_i^{l}$ for $v_i$ by aggregating the information from its neighbors along different relations.
We utilize its neighbors' context information vectors obtained at layer $l-1$, and the computation is as below
\begin{equation}\label{eq:civ}
    \setlength\abovedisplayskip{1pt}
    \setlength\belowdisplayskip{1pt}
    \resizebox{\columnwidth}{!}{$
    c_i^{l} = W_2^{l}\Big(
        ||_{h\in[1,H]} \sigma(
                W_1^{l} \sum\limits_{r_{ST} \in \mathcal{R}} \alpha_{ST}^{h,l} \sum\limits_{v_j \in N_S(i)} c_j^{l-1} + B_1^{l}
            ) + B_2^{l}
        \Big),
    $}
\end{equation}
where $N_S(i)$ denotes the adjacent neighbors of $v_i$ in graph $G_S$ for each relation $r_{ST} \in \mathcal{R}$ relevant to node type $T$, $W_1^l$, $W_2^l$, $B_1^l$, and $B_2^l$ are the trainable parameters in the $l$-th layer, and $H$ is the number of different heads. Note that finally we only use the embedding of the primary graph nodes at the $k$-th layer to obtain $k$-length context information vectors $C_P^k$.

In order to get the information of $k$-length context paths for nodes, we run CP-GNN layer $k$ times to obtain the $C_P^k$ at layer $k$. The GRU mechanism \cite{cho2014learning} is also utilized to alleviate the over smoothing problem unusually occured in GNN model \cite{chen2020measuring}. The computation is as below
\begin{equation}
    \setlength\abovedisplayskip{1pt}
    \setlength\belowdisplayskip{1pt}
    C_P^{l} = GRU\Big(C_P^{l-1}, CP-GNNLayer(C_P^{l-1},C_A^{l-1})\Big),
\end{equation}
where $C_A^{l-1}$ is the embedding of the auxiliary graphs in layer $l-1$, and CP-GNNLayer is the computing process as shown in Eq.~\ref{eq:rlagg} and~\ref{eq:civ}.
Therefore, the final context information vector $c_i^k$ of each node in $G_P$ can be taken from $C_P^k$. The overall process of CP-GNN in shown in Algorithm \ref{alg:CP-GNN}.

\begin{algorithm}[htbp]
    \small
    \caption{The overall process of CP-GNN.}\label{alg:CP-GNN}
    \KwIn {$G=\{\mathcal{V},\mathcal{E},\mathcal{A},\mathcal{R}\}$, $G_P=\{\mathcal{V}_P,\mathcal{E}_P$\}, $K$.}
    \KwOut {The final embedding $Z_P$.}
        Randomly initialize the $Z_P$, and set the loss $L\gets 0$\;
        Initialize relation attention weight $\alpha_1\gets1$,$\cdots$, $\alpha_K\gets1$\;
        \For {$k=1,\cdots,K$}{
            \For{ $v_i \in \mathcal{V}_P$}{
                \For {$v_j \in N_P^k(v_i)$}{
                    $C_P^0\leftarrow Z_P$\;
                    $C_A^0=EmbTransform(Z_P^0)$\;
                    \For {$l=1,\cdots,k$}{
                        $C_P^l=GRU\Big(C_P^{l-1},CP-GNNLayer(C_P^{l-1},C_A^{l-1})\Big)$\;
                    }
                    $L\gets L-\alpha_k \cdot log\varphi(z_i,z_j,c_i^k,c_j^k)$, where $c_i^k,c_j^k\in C_P^k$\;
                }
            }
        }
        Back propagation and update CP-GNN parameters, $\alpha_1,\cdots,\alpha_K$, $Z_P$\;
    \Return {$Z_P$}
\end{algorithm}

\section{Experiments}
\label{sec:exp-first}


\subsection{Dataset}

Three widely used real-world heterogeneous graph datasets, i.e., ACM \cite{wang2019heterogeneous}, DBLP \cite{gao2009graph}, IMDB \cite{wu2016explaining}, together with a schema-rich knowledge graph dataset AIFB \cite{ristoski2016collection} are chosen in the experiments to evaluate the performance of CP-GNN and other baseline models. We report their statistics in Table \ref{tab:dataset}, and discuss their details as follows.

\begin{itemize}
    \item ACM dataset \cite{wang2019heterogeneous} is a bibliographic information network with four types of nodes. We use paper nodes to generate the primary graph and the rest three types of nodes are respectively used to construct auxiliary graphs. 
    In the original dataset, the paper nodes are categorized into 3 classes, i.e., \textit{database}, \textit{wireless communication} and \textit{data mining}. 
    To evaluate the compared models, we pre-define two meta-paths according to \cite{wang2019heterogeneous}, i.e., Paper-Author-Paper (PAP), and Paper-Subject-Paper (PSP).

    \item DBLP dataset \cite{gao2009graph} is a monthly updated citation network consisting of four node types. As we mentioned in Definition \ref{def:ptype}, any node type can be chosen as the primary type. Thus we choose the author node and paper as the primary node, respectively, with four classes, i.e., \textit{database}, \textit{data mining}, \textit{information retrieval} and \textit{machine learning}. 
    They are denoted as DBLP-A and DBLP-P in the following section. We use three meta-paths for this dataset, which are Author-Paper-Author (APA), Author-Paper-Conference-Paper-Author (APCPA), and Author-Paper-Term-Paper-Author (APTPA).

    \item IMDB dataset \cite{wu2016explaining}  consists of four types of nodes, i.e., ``Director'', ``Actors'', ``Movie'' and ``Key word''. We choose the movie node as the primary node with three classes, i.e., \textit{Action}, \textit{Comedy} and \textit{Drama}. 
    We also use three meta-paths on this network, i.e.,  Movie-Actor-Movie (MAM), Movie-Director-Movie (MDM), and Movie-Keyword-Movie (MKM).

    \item AIFB dataset \cite{ristoski2016collection} is a knowledge graph dataset consists of 7 types of nodes and 104 types of edges. We choose the ``Personen'' node as the primary node with four classes.  Due to the complexity of the graph, we do not provide detail illustration in Table \ref{tab:dataset}, and pre-define the meta-paths by ourselves.
\end{itemize}

Notabally, we adopt the meta-paths in ACM, DBLP, and IMDP defined by previous works to evaluate the meta-path-based methods. Due to the rich-schema property of AIFB, we do not define the meta-path by ourselves. Besides, since the unsupervised methods do not need the training data, to make a fair comparison between the unsupervised and supervised methods, the splits of the labled primary nodes for training and testing are shown in Table \ref{tab:datasplit}.

\begin{table}[tbp]
    \caption{Statistics of datasets (the primary node types are marked with ``*''). }
    \label{tab:dataset}
    \Huge
    \resizebox{\columnwidth}{!}{
        \begin{tabular}{@{}cccccc@{}}
            \toprule
            Dataset & Node type                          & \# Nodes    & Edge type           & \# Edges     & Meta-path \\
            \midrule
            ACM     & \tabincell{c}{*Paper (\textbf{P})                                                                 \\ Author (\textbf{A})\\ Subject (\textbf{S}) \\ Facility (\textbf{F})} & \tabincell{c}{12,499\\ 17,431\\ 73 \\ 1,804} & \tabincell{c}{Paper - Paper \\ Paper - Author \\ Paper - Subject \\ Author - Facility} & \tabincell{c}{30,789 \\ 37,055 \\ 12,499 \\ 30,424} & \tabincell{c}{PAP \\ PSP} \\
            \midrule
            DBLP    & \tabincell{c}{*Author (\textbf{A})                                                                \\ *Paper (\textbf{P}) \\ Conference (\textbf{C}) \\ Term (\textbf{T}) } & \tabincell{c}{14,475 \\ 14,736 \\ 20 \\ 8,920} & \tabincell{c}{Author - Paper \\ Paper - Conference \\ Paper - Term} & \tabincell{c}{41,794\\ 14,736 \\ 114,624} & \tabincell{c}{APA \\ APCPA \\ APTPA} \\
            \midrule
            IMDB    & \tabincell{c}{*Movie (\textbf{M})                                                                 \\ Actor (\textbf{A})\\ Director (\textbf{D}) \\ Keyword (\textbf{K})} & \tabincell{c}{4,275 \\ 5,432\\ 2,083 \\ 7,313} & \tabincell{c}{Movie - Actor \\ Movie - Director \\ Movie - Keyword} & \tabincell{c}{12,831 \\ 4,181 \\ 20,428} & \tabincell{c}{MAM \\ MDM \\ MKM} \\
            \midrule
            AIFB    & 7 different types                  & Total 7,262 & 104 different types & Total 48,810 & -         \\
            \bottomrule
        \end{tabular}
    }
\end{table}

\begin{table}[htbp]
    \caption{Statistics of the training and testing set.}
    \centering
    \label{tab:datasplit}
    \resizebox{0.65\columnwidth}{!}{
    \begin{tabular}{@{}cccc@{}}
        \toprule
        Dataset & Primary Type & Training & Testing \\ \midrule
        ACM     & Paper        & 805      & 3,220   \\
        DBLP-A  & Author       & 811      & 3,246   \\
        DBLP-P  & Paper        & 20       & 80      \\
        IMDB    & Movie        & 855      & 3,420   \\
        AIFB    & Personen     & 36       & 141     \\ \bottomrule
    \end{tabular}
    }
\end{table}

\subsection{Baseline Methods}
To evaluate the effectiveness of our approach, we compare it with a list of the state-of-the-art unsupervised methods (i.e., Node2vec \cite{grover2016node2vec}, Metapath2vec \cite{dong2017metapath2vec}, and HIN2vec \cite{fu2017hin2vec}) and supervised baseline methods (i.e., GCN \cite{kipf2016semi}, GAT \cite{velivckovic2017graph}, LGNN \cite{chen2019supervised}, HAN \cite{wang2019heterogeneous}, and HGT \cite{hu2020heterogeneous}). Especially, HGT is a semi-supervised neural network model which adopts the transformer mechanism to capture the importance of relations. It is considered as the SOTA approach.

Except for graph embedding-based and GNN-based baselines, we also choose two traditional community detection methods (i.e., InfoMap \cite{rosvall2008maps} and LP (Label Propagation) \cite{raghavan2007near}), for comparison. 

\subsection{Settings of Model Parameters}
We now briefly discuss the settings of model parameters.
For unsupervised approaches such as Node2vec, Metapath2vec, and HIN2vec, we respectively set the length of a random walk to 20, the sampling window size to 3, the number of walks per node to 5, and the number of negative samplings to 3.
For supervised-based methods such as GCN, GAT, LGNN, HAN and HGT, the number of graph convolution layer is set to 2, and their node features are first randomly initialized then updated during the model learning process. The dimension of node feature embedding for all compared methods is set to 128. 

For our CP-GNN, the number of attention heads is set to 8, the dimension of the output vectors of K/Q-Linear components is set to 128, and the node dropout rate is set to 0.3. The number of positive context neighbors for each node in a context path is set to 20, and the corresponding negative sampling size is set to 3. The Adam \cite{kingma2014adam} is adopted to optimize all models, and the learning rate is set to 0.05.

\begin{table*}[htbp]
    \centering
    \caption{Comparison of the performance of different community detection methods.}
    \label{tab:cl_result}
    \resizebox{0.83\textwidth}{!}{
    \begin{tabular}{@{}ccccccccccccc@{}}
        \toprule
        Dataset                 & Metrics & InfoMap & Node2vec & Metapath2vec & HIN2vec & LP      & GCN     & GAT     & LGNN    & HAN            & HGT             & CP-GNN          \\ \midrule
        \multirow{4}{*}{ACM}    & F1      & 0.5733  & 0.6954   & 0.7142       & 0.7732  & 0.6691  & 0.5366  & 0.6876  & 0.6987  & 0.7922         & 0.7599          & \textbf{0.8596} \\
                                & NMI     & 0.1933  & 0.2666   & 0.3596       & 0.4066  & 0.3933  & 0.0966  & 0.2577  & 0.2746  & 0.394          & 0.4509          & \textbf{0.4832} \\
                                & ARI     & 0.1286  & 0.2469   & 0.2956       & 0.3313  & 0.2992  & 0.1022  & 0.1422  & 0.2368  & 0.319          & 0.3813          & \textbf{0.3924} \\
                                & Purity  & 0.5876  & 0.4355   & 0.4969       & 0.6969  & 0.6691  & 0.5808  & 0.6186  & 0.6594  & 0.6942         & 0.7032          & \textbf{0.715}  \\\midrule
        \multirow{4}{*}{DBLP-A} & F1      & 0.3601  & 0.7572   & 0.7144       & 0.313   & 0.2451  & 0.32    & 0.9023  & 0.321   & 0.9023         & \textbf{0.9386} & 0.9125          \\
                                & NMI     & 0.0819  & 0.0638   & 0.2554       & 0.0044  & 0.0984  & 0.0186  & 0.618   & 0.0069  & 0.624          & 0.7032          & \textbf{0.7089} \\
                                & ARI     & 0.0131  & 0.0409   & 0.2722       & 0.0022  & 0.0033  & 0.0166  & 0.5264  & -0.0012 & 0.665          & 0.7322          & \textbf{0.766}  \\
                                & Purity  & 0.3601  & 0.3884   & 0.6169       & 0.2971  & 0.2935  & 0.3564  & 0.7476  & 0.2988  & 0.8496         & \textbf{0.9325} & 0.9004          \\\midrule
        \multirow{4}{*}{DBLP-P} & F1      & 0.3111  & 0.3      & 0.3125       & 0.3375  & 0.4     & 0.31    & 0.3     & 0.225   & 0.3375         & 0.4             & \textbf{0.4875} \\
                                & NMI     & 0.0463  & 0.0655   & 0.0034       & 0.0514  & 0.0429  & 0.0171  & 0.0495  & 0.0431  & 0.0732         & 0.1086          & \textbf{0.1846} \\
                                & ARI     & 0.0032  & -0.0016  & 0.0013       & -0.0021 & 0.0042  & -0.0048 & -0.0029 & 0.0016  & -0.0103        & 0.0724          & \textbf{0.0564} \\
                                & Purity  & 0.4111  & 0.432    & 0.4212       & 0.431   & 0.4     & 0.41    & 0.422   & 0.44    & 0.426          & 0.426           & \textbf{0.488}  \\\midrule
        \multirow{4}{*}{IMDB}   & F1      & 0.3038  & 0.5494   & 0.488        & 0.4184  & 0.3826  & 0.3628  & 0.3587  & 0.3646  & 0.4888         & 0.3634          & \textbf{0.614}  \\
                                & NMI     & 0.0098  & 0.0745   & 0.027        & 0.0031  & 0.0081  & 0.0018  & 0.0012  & 0.0158  & 0.1172         & 0.0101          & \textbf{0.1225} \\
                                & ARI     & -0.005  & 0.0471   & 0.0146       & -0.0022 & -0.0004 & 0.0013  & -0.0009 & -0.0079 & \textbf{0.131} & 0.0083          & 0.1231          \\
                                & Purity  & 0.3898  & 0.4442   & 0.438        & 0.3744  & 0.3825  & 0.3885  & 0.3746  & 0.3738  & 0.3734         & 0.4023          & \textbf{0.4949} \\\midrule
        \multirow{4}{*}{AIFB}   & F1      & 0.434   & 0.7517   & -            & 0.6524  & 0.4151  & 0.6524  & 0.7375  & 0.6809  & -              & 0.7163          & \textbf{0.7659} \\
                                & NMI     & 0.0645  & 0.2401   & -            & 0.1912  & 0.2216  & 0.1567  & 0.2117  & 0.2435  & -              & 0.3812          & \textbf{0.4147} \\
                                & ARI     & 0.0286  & 0.1518   & -            & 0.1202  & 0.0985  & 0.1248  & 0.1142  & 0.079   & -              & 0.3011          & \textbf{0.3898} \\
                                & Purity  & 0.4403  & 0.6091   & -            & 0.5494  & 0.5157  & 0.5835  & 0.5568  & 0.5875  & -              & \textbf{0.7102} & 0.6966          \\ \bottomrule
    \end{tabular}
    }
\end{table*}

\subsection{Performance Comparison}
\label{sec:expRQ1}

In this experiment, we first learn the node embeddings and then employ the $k$-Means algorithm on these embeddings to detect communities, where $k$ is set to the number of node classes.
We evaluate the community detection results using the ground-truth labels with four commonly adopted community detection evaluation metrics, i.e., F1, NMI, ARI, and Purity, and report the performance results in Table \ref{tab:cl_result}. Note that due to the lack of pre-defined meta-paths, the meta-path-based methods ( i.e., Metapath2vec and HAN) cannot be evaluated on AIFB.

From Table \ref{tab:cl_result}, we can clearly see that CP-GNN outperforms all the baselines on most metrics. This demonstrates that via the context paths, it can learn node representations that are more suitable for community detection by capturing more meaningful and high-order relationships.

For unsupervised methods, HIN2vec achieves better performance than Node2vec and Metapath2vec on ACM but worse on other datasets. It shows that even though HIN2vec can automatically discover meta-paths by random walk, the discovered meta-paths may not be suitable for community detection. Besides, HIN2vec does not differentiate the importance of the found meta-paths, which incorporates some unrelated relations to the community detection result. The performances of the Metapath2vec are better on DBLP than ACM, when using the longer meta-path. This demonstrates the importance of high-order relationships. Last, all the graph embedding-based methods are better than InfoMap, which shows the great potential of them in community detection task.

For supervised methods, LP performs well in schema-simple heterogeneous graphs (i.e., ACM and IMDB), but its performance drops quickly when it meets the schema-rich heterogeneous graph (i.e., AIFB). This indicates the simple label propagation mechanism does not consider the complex relations in heterogeneous graphs, which leads to its poor performance. GCN and LGNN achieve the worst results in the GNN-based baselines. The possible reason is that they are originally proposed for homogeneous graph, thus they do not consider the complex context information in heterogeneous graph. GAT performs better than GCN and LGNN, which strongly supports the importance of the attention mechanism. The attention mechanism used in GAT can be regarded as a simple way to differentiate the node type and edge type in heterogeneous graph. Thanks to the meta-path, HAN can explicitly excavates the complex semantic information and reaches a better result. In addition, HGT achieves the second-best performance since it can capture more diverse relations without the limitation of the meta-paths, and be easily fit into different datasets. However, the HGT still requires the labeled data to optimize the model, which strongly limits its application.

Last, as Definition \ref{def:ptype} said, each node type can be treated as primary type. Therefore, the result in DBLP-A and DBLP-P shows that no matter what node type is chosen as the primary type, the proposed GP-GNN can still achieve better results with the help of other auxiliary nodes.

\subsection{Parameters Analysis and Ablation Study}\label{sec:parameter}
\label{sec:expRQ2}

\begin{figure}[htbp]
    \centering
    \setlength{\abovecaptionskip}{1pt}
    \setlength{\belowcaptionskip}{1pt}
    \includegraphics[trim={5 5 5 0},width=0.8\columnwidth,clip]{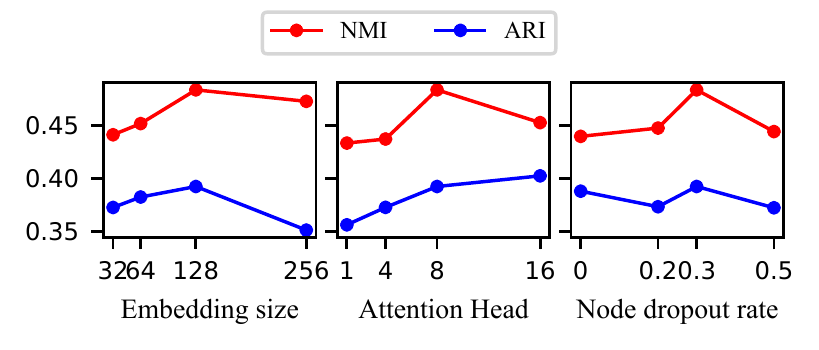}
    \caption{Parameter sensitivity w.r.t. different parameters.}
    \label{fig:parameter}
\end{figure}

In this section, we experimentally investigate the sensitivity of the parameters and report the results on the ACM dataset with various parameters shown in Figure \ref{fig:parameter}, and various CP-GNN structures shown in Tables \ref{tab:tranfunc} and \ref{tab:attenmethods}.

\begin{table}[htbp]
    \caption{Effectiveness of embedding transform functions.}
    \centering
    \label{tab:tranfunc}
    \begin{tabular}{@{}ccc@{}}
        \toprule
        Trans.   Function   & NMI             & ARI             \\ \midrule
        CP-GNN $w/o$ Trans. & 0.3397          & 0.3665          \\
        CP-GNN $w/o$ ReLU   & 0.3908          & 0.2908          \\
        CP-GNN              & \textbf{0.4832} & \textbf{0.3924} \\ \bottomrule
    \end{tabular}
\end{table}

As shown in Figure \ref{fig:parameter}, with the increase of parameter values, CP-GNN's performances raise first and then drop slightly; the best performances are reached when the Embedding size, Attention head, and Node droupout rate reach 128, 8, and 0.3, respectively. The reason is that an over large embedding dimension may introduce unnecessary redundancies to the CP-GNN model. Although more attention heads can capture more diverse relation importance and increase the representation ability, they also introduce more parameters to the model, making it hard to train. Besides, too many nodes are dropped, which causes that the graph summary vectors cannot be well generated from the remaining nodes.

In Table \ref{tab:tranfunc}, we evaluate the effect of the embedding transformation mechanism where \textit{CP-GNN $w/o$ Trans.} means that the initial embeddings of all the auxiliary nodes are randomly initialized and optimized during the training; \textit{CP-GNN $w/o$ ReLU} denotes that the initial embeddings are transformed from the primary node embeddings but without non-linear function in Eq.~\ref{eq:transfunc}. \textit{CP-GNN} denotes the final embedding transformation function used in our CP-GNN.
The experiment results show that our non-linear transformation function performs the best. The reason is that using the proposed transformation function can establish stronger connections between the primary graph and auxiliary graphs. Meanwhile, it provides different contextual representation of the primary graph. Besides, the non-linear function, such as ReLU, can ``mask'' some unimportant features during the transformation to provide better result.

In Table \ref{tab:attenmethods}, to examine the effectiveness of different attention methods, we gradually remove the relation and length attention mechanism. \textit{CP-GNN $w/o$ Attention} means CP-GNN without any attention mechanism. \textit{CP-GNN $w/o$ Relation Att.} and \textit{CP-GNN $w/o$ Att.} respectively denote the CP-GNN without relation attention and context path length attentnion. The experimental results show that both the context path length attention and the relation attention are helpful for improving the model performance.

\begin{table}[tbp]
    \caption{Effectiveness of attention mechanisms.}
    \centering
    \label{tab:attenmethods}
    \resizebox{0.6\columnwidth}{!}{
    \begin{tabular}{@{}ccc@{}}
        \toprule
        Attention                  & NMI             & ARI             \\\midrule
        CP-GNN $w/o$ Attention     & 0.2214          & 0.1132          \\
        CP-GNN $w/o$ Relation Att. & 0.3239          & 0.1008          \\
        CP-GNN $w/o$ Length Att.   & 0.4452          & 0.3739          \\
        CP-GNN                     & \textbf{0.4832} & \textbf{0.3924} \\
        \bottomrule
    \end{tabular}
    }
\end{table}

\subsection{Case Study}

\subsubsection{Context Path Length Attention}
\begin{figure}[htbp]
    \centering
    \setlength{\abovecaptionskip}{1pt}
    \setlength{\belowcaptionskip}{1pt}
    \begin{subfigure}[t]{0.48\columnwidth}
        \includegraphics[width=1.\linewidth]{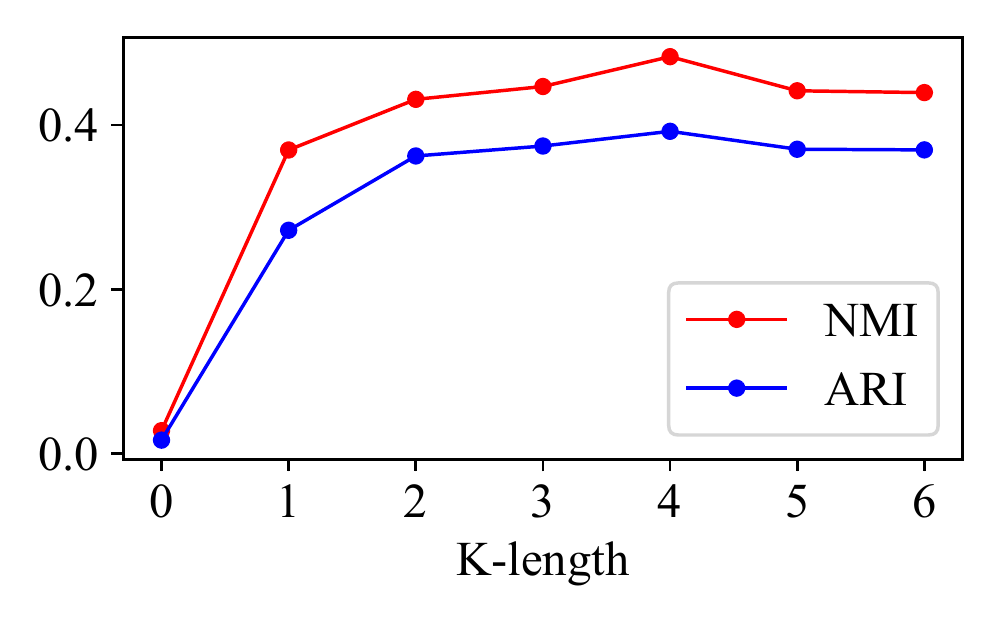}
        \caption{CP-GNN's performances under different $K$-length context paths.}
        \label{fig:k-length}
    \end{subfigure}
    \begin{subfigure}[t]{0.48\columnwidth}
        \includegraphics[width=1.\linewidth]{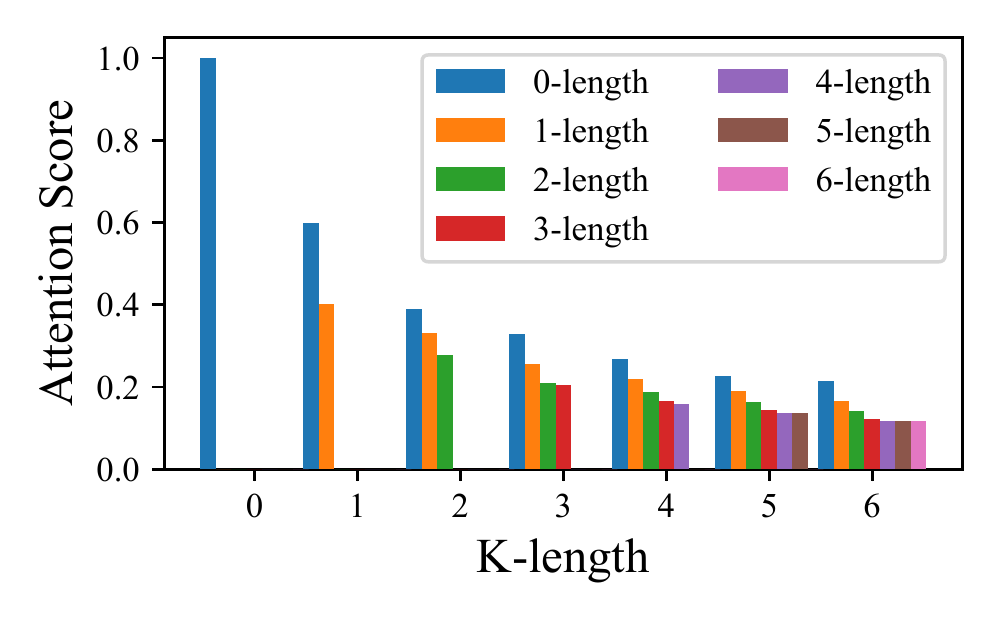}
        \caption{Attention weight of each context path length.}
        \label{fig:k-attention}
    \end{subfigure}
    \caption{Visualization of the context path length attention mechanism.}
    \label{fig:atten_matrix}
\end{figure}

To analyze the effect of context path length attention, we conduct experiments on the ACM dataset with different maximum context path length $K$. The experiment results and attention weight of each context path length are depicted in Figure \ref{fig:k-length} and \ref{fig:k-attention}.
Clearly, with the increase of $K$, the performances of CP-GNN increases and reaches the best when $K$=4. This demonstrates our assumption that the high-order relationship information is crucial for community detection. When $K>4$, the performance of the model slightly drops, which may due to the fact that the nodes will be fully connected when $K$ is overly increasing.

Besides, by reporting the attention score of each length, we can see that with the context path length increases, the corresponding attention weight decreases, which is consistent to the common sense that the short paths often reflect stronger connections than the long ones. What is more, the attention scores of lengths longer than 4 (i.e., $k=5,6$) are barely the same. It indicates the less relevance of these paths, and explains why the performance of CP-GNN drops slightly when $K>4$.
By using the context path length attention, we not only capture the high-order relationships among the graph nodes, but also discard some less important high-order relationships during learning.

Since the context path length attention can discriminate the importance of context paths with different lengths and assign lower importance to unnecessary long meta-paths, we can set a slightly larger value for $K$, which can yield a result closed to the optimal one. As demonstrated by Figure \ref{fig:k-length} where the optimal result is reached when $K$=4 but the results are close when $K$=5 or 6. This will alleviate the model's dependence on the hyper-parameter $K$.

\begin{figure}[htbp]
    \centering
    \setlength{\abovecaptionskip}{1pt}
    \setlength{\belowcaptionskip}{1pt}
    \begin{subfigure}[b]{0.54\columnwidth}
        \includegraphics[width=1.\linewidth]{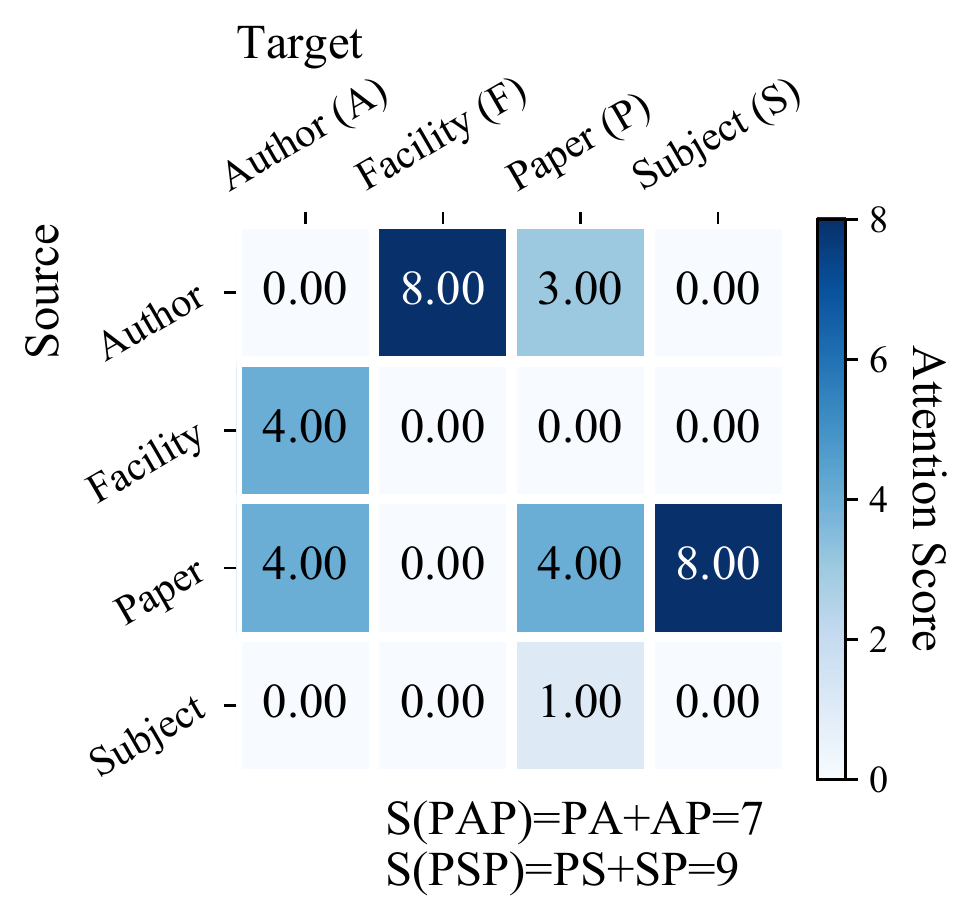}
        \caption{Attention matrix of 1-length context path on ACM.}
        \label{fig:ACMMatrix}
    \end{subfigure}
    \begin{subfigure}[b]{0.45\columnwidth}
        \includegraphics[width=1.\linewidth]{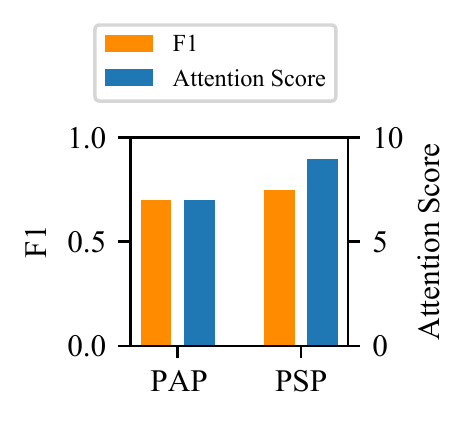}
        \caption{Metapath2vec F1 values on ACM.}
        \label{fig:ACMResult}
    \end{subfigure}
    \caption{Visualization of the context relation attention matrix on ACM.}
    \label{fig:rel_atten_matrix}
\end{figure}

\begin{figure}[tbp]
    \centering
    \setlength{\abovecaptionskip}{1pt}
    \setlength{\belowcaptionskip}{1pt}
    \begin{subfigure}[b]{0.54\columnwidth}
        \includegraphics[width=1.\linewidth]{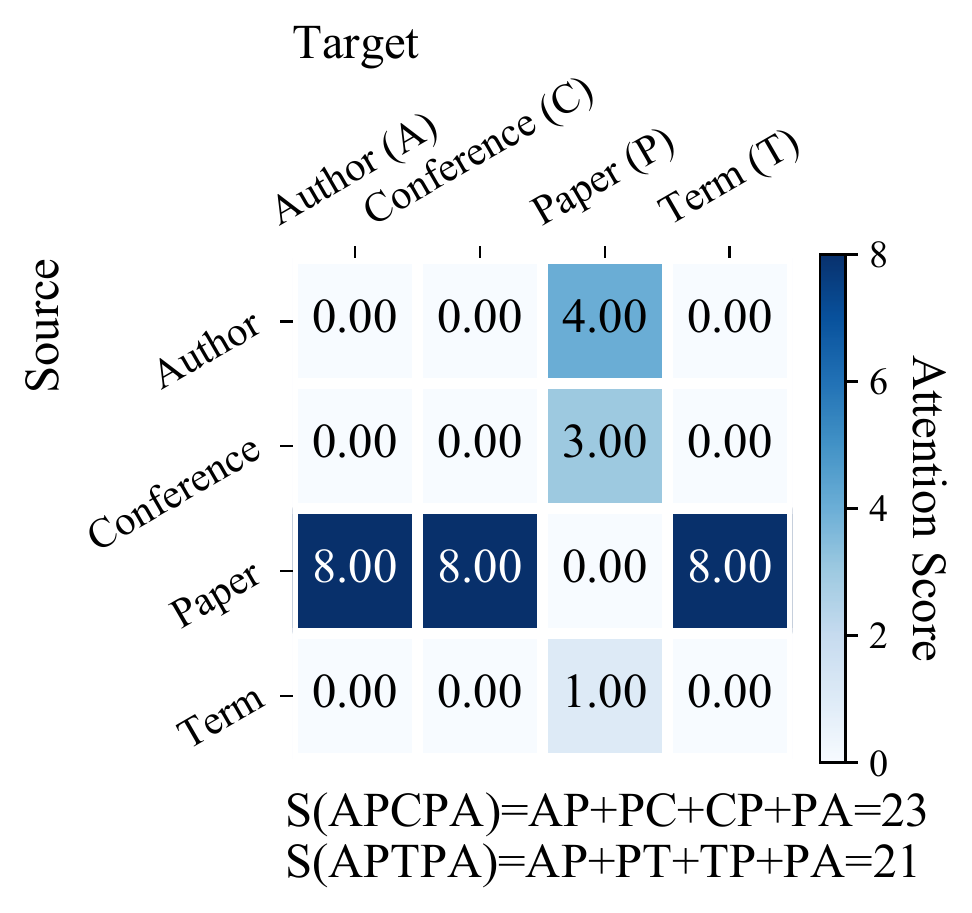}
        \caption{Attention matrix of 3-length context path on DBLP.}
        \label{fig:DBLPMatrix}
    \end{subfigure}
    \begin{subfigure}[b]{0.45\columnwidth}
        \includegraphics[width=1.\linewidth]{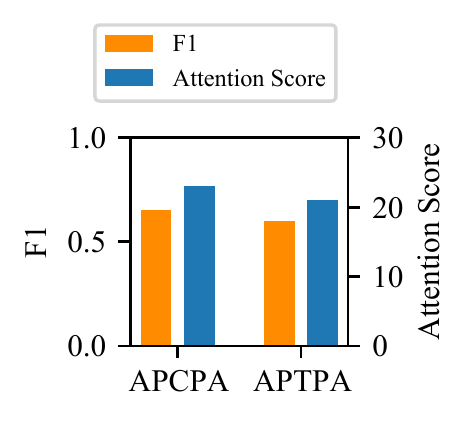}
        \caption{Metapath2vec F1 values on DBLP.}
        \label{fig:DBLPResult}
    \end{subfigure}
        \caption{Visualization of the context relation attention matrix on DBLP.}
    \label{fig:DBLP_rel_atten_matrix}
\end{figure}

\subsubsection{Relation Attention}

To further analyze whether CP-GNN can differentiate the context paths, we first present the corresponding relations attention matrix acquired from CP-GNN in Figure \ref{fig:ACMMatrix} and \ref{fig:DBLPMatrix}, where each entry is the attention score of the relation with a source node type and a target node type. The attention score of each context path can be computed by summarizing the scores of its relations. Then, to justify whether the paths with higher attention score are more meaningful for community detection, we adopt the Metapath2vec to evaluate the effect of each path. The results are shown in Figure \ref{fig:ACMResult} and \ref{fig:DBLPResult}

For example, the paths PAP and PSP in ACM are both 1-length context path. Therefore, there attention scores can be computed from the relation attention matrix of 1-length context path shown in Figure \ref{fig:ACMMatrix} where $S(PAP)=PA+AP=7$ and $S(PSP)=PS+SP=9$. Clearly, we can find that the attention score of PSP is higher than PAP, which means the path PSP is slightly more important than PAP for community detection. This can be justified by the result shown in Figure \ref{fig:ACMResult} where the F1 score of PSP is higher than PAP.
Similarly, from Figure \ref{fig:DBLPMatrix}, the attention scores of paths APCPA and APTPA are $S(APCPA)=AP+PC+CP+PA=23$ and $S(APTPA)=AP+PT+TP+PA=21$. This indicates that the relationship reflected by APCPA is a little bit more important than that of APTPA. This finding can be proved by the result shown in Figure \ref{fig:DBLPResult} where the F1 of APCPA is higher than APTAP.

In summary, the above analysis demonstrates that the Relation Attention can discover context path of different importance and capture the context path of high importance that are more meaningful and useful for community detection.

\section{Conclusion}
\label{sec:conclude}

In this paper, we propose the Context Path-based Graph Neural Network (CP-GNN) model for detecting communities from heterogeneous graphs, which not only avoids using pre-defined meta-paths, but also well  captures the high-order relationship among nodes.
In particular, we adopt the context path and propose the context path probability to model the objective function. Besides, CP-GNN distinguishes the importance of different context paths.
Extensive experiments on real-world datasets show that CP-GNN outperforms the baselines, and its attention mechanisms can well differentiate the importance of different context paths. 

\section{Acknowledgments}
This work is supported in part by the National Natural Science Foundation of China under Grant No. 61872108, and the Shenzhen Science and Technology Program under Grant No. JCYJ2020010911 3201726, JCYJ20170811153507788. Xin Cao is supported by ARC DE190100663. Yixiang Fang is supported by CUHK-SZ grant UDF0 1002139. Wenjie Zhang is supported by ARC DP200101116.

\clearpage

\bibliographystyle{ACM-Reference-Format}
\bibliography{sections/myref}










\end{document}


\maketitle

\section{Dataset}

Three widely used real-world heterogeneous graph datasets, i.e., ACM \cite{wang2019heterogeneous}, DBLP \cite{gao2009graph}, IMDB \cite{wu2016explaining}, together with a schema-rich knowledge graph dataset AIFB \cite{ristoski2016collection} are chosen in the experiments to evaluate the performance of CP-GNN and other baseline models. We report their statistics in Table \ref{tab:dataset}, and discuss their details as follows.

\begin{itemize}
    \item ACM dataset \cite{wang2019heterogeneous} is a bibliographic information network with four types of nodes. We use paper nodes to generate the primary graph and the rest three types of nodes are respectively used to construct auxiliary graphs. In the original dataset, the paper nodes are categorized into 3 classes, i.e., \textit{database}, \textit{wireless communication} and \textit{data Mining}. To evaluate the compared models, we pre-define two meta-paths according to \cite{wang2019heterogeneous}, i.e., Paper-Author-Paper (PAP), and Paper-Subject-Paper (PSP).
    
    \item DBLP dataset \cite{gao2009graph} is a monthly updated citation network consisting of four node types, and we choose the author node as the primary node with four classes, i.e., \textit{database}, \textit{data mining}, \textit{information retrieval} and \textit{machine learning}.
    We use three meta-paths this dataset, which are Author-Paper-Author (APA), Author-Paper-Conference-Paper-Author (APCPA), and Author-Paper-Term-Paper-Author (APTPA).
    
    \item IMDB dataset \cite{wu2016explaining}  consists of three types of nodes, i.e., ``director'', ``actors'' and ``Movie Release Date''. We choose the movie node as the primary node with three classes, i.e., \textit{Action}, \textit{Comedy} and \textit{Drama}. We also use three meta-paths on this network, i.e.,  Movie-Actor-Movie (MAM), Movie-Director-Movie (MDM), and Movie-Keyword-Movie (MKM).
    \item AIFB dataset \cite{ristoski2016collection} is a knowledge graph dataset consists of 7 types of nodes and 104 types of edges. We choose the ``Personen'' node as the primary node with four classes.  Due to the complexity of the graph, we do not provide detail illustration in Table \ref{tab:dataset}, and pre-define the meta-paths by ourselves.
\end{itemize}

\begin{table}[htbp]
    \caption{Statistics of datasets (the primary node types are marked with ``*''). }
    \label{tab:dataset}
    \Huge
    \resizebox{\columnwidth}{!}{
        \begin{tabular}{@{}c|cccc|c@{}}
            \toprule
            Dataset & Node type                          & \# Nodes   & Edge type           & \# Edges    & Meta-path \\
            \midrule
            ACM     & \tabincell{c}{*Paper (\textbf{P})                                                               \\ Author (\textbf{A})\\ Subject (\textbf{S}) \\ Facility (\textbf{F})} & \tabincell{c}{12499\\ 17431\\ 73 \\ 1804} & \tabincell{c}{Paper - Paper \\ Paper - Author \\ Paper - Subject \\ Author - Facility} & \tabincell{c}{30789 \\ 37055 \\ 12499 \\ 30424} & \tabincell{c}{PAP \\ PSP} \\
            \midrule
            DBLP    & \tabincell{c}{*Author (\textbf{A})                                                              \\ Paper (\textbf{P}) \\ Conference (\textbf{C}) \\ Term (\textbf{T}) } & \tabincell{c}{14475 \\ 14736 \\ 20 \\ 8920} & \tabincell{c}{Author - Paper \\ Paper - Conference \\ Paper - Term} & \tabincell{c}{41794\\ 14736 \\ 114624} & \tabincell{c}{APA \\ APCPA \\ APTPA} \\
            \midrule
            IMDB    & \tabincell{c}{*Movie (\textbf{M})                                                               \\ Actor (\textbf{A})\\ Director (\textbf{D}) \\ Keyword (\textbf{K})} & \tabincell{c}{4275 \\ 5432\\ 2083 \\ 7313} & \tabincell{c}{Movie - Actor \\ Movie - Director \\ Movie - Keyword} & \tabincell{c}{12831 \\ 4181 \\ 20428} & \tabincell{c}{MAM \\ MDM \\ MKM} \\
            \midrule
            AIFB    & 7 different types                  & Total 7262 & 104 different types & Total 48810 & -         \\
            \bottomrule
        \end{tabular}
    }
\end{table}

\section{Baseline Methods}
\label{sec:baseline}

To evaluate the effectiveness of our approach, we compare it with the state-of-the-art unsupervised methods and supervised baseline methods.

\subsection{Unsupervised Baseline Methods}
\begin{itemize}
    \item Node2vec \cite{grover2016node2vec} is a random walk-based graph embedding method originally proposed for analyzing homogeneous graphs. For simplicity, we ignore the heterogeneous node types and edge types in the experiments.
    \item Metapath2vec \cite{dong2017metapath2vec} is a heterogeneous graph embedding method which performs meta-path-based random walk and employs the skip-gram model to embed each node. 
    \item HIN2vec \cite{fu2017hin2vec} is a unsupervised heterogeneous graph embedding method which can automatically find the useful meta-path from the heterogeneous graph without pre-defining them.
\end{itemize}

\subsection{Supervised Baseline Methods}
\begin{itemize}
    \item GCN \cite{kipf2016semi} is considered as a benchmark graph convolutional network model, originally proposed for semi-supervised classification on homogeneous graph. For simplicity, we ignore the types of nodes and edges in the experiments.

    \item GAT \cite{velivckovic2017graph} is a GNN model that uses a hidden self-attention layer to assign different weights to different node features. Similarly, we ignore the type of nodes and edges.

    \item LGNN \cite{chen2019supervised} is GNN model that transform the graph into line graph and perform supervised community detection task on it.

    \item HAN \cite{wang2019heterogeneous} models the heterogeneous graph by discovering both node-level and semantic-level information via designed attention mechanism.

    \item HGT \cite{hu2020heterogeneous} is a semi-supervised neural network model which adopts the transformer mechanism to capture the importance of relations which is considered as the SOTA approach.
\end{itemize}

\section{Settings of Model Parameters}
\label{sec:expsettings}

We now briefly discuss the settings of model parameters.
For unsupervised approaches such as Node2vec, Metapath2vec, and HIN2vec, we respectively set the length of a random walk to 20, the sampling window hop size to 3, the number of walks per node to 5, and the number of negative samplings to 3.
For supervised-based methods such as GCN, GAT, LGNN, HAN and HGT, the dimension of node feature embedding for all compared methods is set to 128. These features are first randomly initialized and then updated during the model learning process.

For our CP-GNN, the number of attention heads is set to 8, the dimension of the output vectors of  K/Q-Linear components is set to 128, and the node dropout rate is set to 0.3. The number of positive context neighbors for each node in a context path is set to 20 and the corresponding negative sampling size is set to 3. The Adam \cite{kingma2014adam} is adopted to optimize all models and the learning rate is set to 0.05.

For a fair comparison between unsupervised methods and supervised methods, we split the primary nodes with 20\% and 80\% for training and testing, respectively.

\bibliographystyle{plain}
\bibliography{sections/myref}